\documentclass[aps,physrev,twocolumn,superscriptaddress,amsmath,amssymb,longbibliography,nofootinbib]{revtex4-2}
\usepackage{amsmath}
\usepackage{amsfonts}
\usepackage{amssymb}
\usepackage{braket}
\usepackage{tcolorbox}
\usepackage{xcolor}
\usepackage{graphicx}
\usepackage{enumerate}
\usepackage{subfigure}
\usepackage{hyperref}

\begin{document}

\title{The anomalously slow dynamics of inhomogeneous quantum annealing}
\author{Mohammadhossein Dadgar}\email{dadgarmo@msu.edu}
\affiliation{Department of Physics and Astronomy, Michigan State University, East Lansing, Michigan 48824, USA}
\author{Christopher L. Baldwin}\email{baldw292@msu.edu}
\affiliation{Department of Physics and Astronomy, Michigan State University, East Lansing, Michigan 48824, USA}

\date{\today}

\begin{abstract}
Inhomogeneous quantum annealing (IQA), in which transverse fields are turned off one by one rather than simultaneously, has been proposed as an effective way to avoid the first-order phase transitions that impede conventional quantum annealing (QA).
Here we explicitly study the dynamics of IQA, rather than merely the thermodynamics, and find that it is appreciably slower than the phase diagram would suggest.
Interestingly, this slowdown manifests both when IQA succeeds in circumventing phase transitions and when it fails.
Even in the absence of transitions, such as for the mean-field models that have been analyzed previously, IQA is slower than expected by a factor of the number of spins $N$.
More significantly, we show that in non-mean-field models, first-order transitions are likely to be quite common, and the gap at such transitions is not merely exponential in $N$ but exactly zero.
Thus IQA cannot reach the ground state on any timescale.
Both of these results can be understood through the simple observation that a spin's magnetization becomes conserved once its field is turned off during the IQA protocol.
\end{abstract}

\maketitle

\section{Introduction} \label{sec:introduction}

Among the many quantum algorithms and protocols that have been proposed, quantum annealing (QA) stands out as a popular method for solving complicated spin-glass and optimization problems~\cite{QuantumAnnealingNew1994finnila, QuantumAnnealingTransverse1998kadowaki, QuantumAnnealingDisordered1999brooke, QuantumAdiabaticEvolution2001farhi, TheoryQuantumAnnealing2002santoro, OptimizationUsingQuantum2006santoro, MathematicalFoundationQuantum2008morita, QuantumAnnealingManufactured2011johnson, EvidenceQuantumAnnealing2014boixo, EntanglementQuantumAnnealing2014lanting, QuantumAnnealingCombinatorial2018kumar, QuantumAnnealingPrime2018jiang, AdiabaticQuantumComputation2018albash, PerspectivesQuantumAnnealing2020hauke, Rajak2023Quantum, EffectivePrimeFactorization2024ding}.
Suppose that the problem can be expressed as finding the ground state of a Hamiltonian $H_0$ involving $N$ spin-1/2s (``qubits'').
We will specifically assume, as is by far the most common case, that $H_0$ is diagonal in the $z$ basis.
Conventional QA applies a uniform transverse field whose strength is slowly reduced from effectively infinite to zero --- the adiabatic theorem~\cite{AdiabaticTheoremQuantum1950kato,Messiah1962,Jansen2007Bounds} guarantees that the spins will remain near the ground state throughout this protocol if the field is varied sufficiently slowly.
Since the initial ground state has all spins aligned with the transverse field and the final ground state is that of $H_0$, this gives a straightforward and elegant means to find the desired state starting from a trivial one.
Unfortunately, it generically requires prohibitively slow annealing rates (namely exponential in $N$) to succeed in hard problems~\cite{SimpleGlassModels2008jorg,Young2010First,Jorg2010Energy,FirstOrderTransitionsPerformance2010jorgb,Hen2011Exponential,QuantumMeanfieldModels2012bapst,Farhi2012Performance}.
In physical terms, the failure is tied to the prevalence of first-order phase transitions in ``rugged energy landscapes''~\cite{Altshuler2010Anderson,Foini2010Solvable,Bapst2013Quantum,Knysh2016ZeroTemperature,Baldwin2018Quantum}.

As a result, researchers have begun exploring modifications to conventional QA, such as incorporating transverse interactions~\cite{ManybodyTransverseInteractions2012seoane, QuantumAnnealingAntiferromagnetic2012seki, QuantumAnnealingAntiferromagnetic2015seki, NonstoquasticHamiltoniansQuantum2017vinci, NonstoquasticHamiltoniansQuantum2017hormozi, ExponentialEnhancementEfficiency2017nishimori, QuantumMonteCarlo2017ohzeki, RelationQuantumFluctuations2017susa}, using non-uniform fields~\cite{Farhi2011Quantum,DoesAdiabaticQuantum2011dickson,AlgorithmicApproachAdiabatic2012dickson}, and reverse annealing~\cite{PerdomoOrtiz2011Study, ReverseAnnealingFully2018ohkuwa, ReverseQuantumAnnealing2019venturelli, TravelTimeOptimization2022haba, CounterdiabaticReverseAnnealing2023passarelli}.
One prominent example is inhomogeneous quantum annealing (IQA)~\cite{ExponentialSpeedupQuantum2018susa, QuantumAnnealing$p$spin2018susa,QuantumPhaseTransition2019hartmann}, in which the transverse fields are reduced to zero one by one rather than simultaneously.
IQA is particularly encouraging because it has been shown to circumvent problematic phase transitions in solvable mean-field models~\cite{ExponentialSpeedupQuantum2018susa} (and similar improvements have been observed in inhomogeneous quenches through quantum critical points~\cite{DynamicsInhomogeneousQuantum2010dziarmaga,InhomogeneousQuasiadiabaticDriving2016rams,Agarwal2018Fast,UniversalDynamicsInhomogeneous2019gomez-ruiz}), suggesting that the annealing can be performed much more rapidly.

In this paper, however, we show that IQA is likely to fail in difficult non-mean-field problems.
Interestingly (albeit unfortunately), the failure is even more extreme than that of conventional QA --- not only will there be phase transitions, but those transitions are associated with exact level crossings and thus IQA cannot reach the ground state on any timescale (even exponential in $N$).
As an additional observation, we also note that even when IQA does avoid phase transitions, the timescale to reach the ground state is still larger by a factor of $N$ compared to conventional QA in the absence of transitions.
Together, these results establish that IQA is \textit{anomalously slow} --- it requires significantly more time than the conventional protocol would under comparable circumstances.
The only hope for IQA (at least in its idealized form) is thus for it to completely remove transitions in a problem for which conventional QA cannot, but this is precisely what we argue below is unlikely to occur in practice.

To be more concrete, conventional QA evolves the spins under the Hamiltonian
\begin{equation} \label{eq:conventional_QA_Hamiltonian}
H(s) = s H_0(\hat{\sigma}^z) - (1-s) \sum_{j=1}^N \hat{\sigma}_j^x,
\end{equation}
where $\hat{\sigma}_j^x$ is the Pauli $x$-operator acting on spin $j$ and $H_0(\hat{\sigma}^z)$ is the target Hamiltonian, which we assume is a function of the Pauli $z$-operators $\{\hat{\sigma}_j^z\}$ alone (and thus is trivially diagonal in that basis).
One begins with all spins pointing in the $x$ direction, and then monotonically increases $s$ from 0 to 1 in time $T$ (it is simplest to take $s(t) = t/T$).
As long as the gap (difference in energy between ground and first excited state) is not identically zero, the adiabatic theorem guarantees that the system will remain near its ground state for sufficiently large $T$.
The question, of course, is how large $T$ has to be, and in particular how it scales with the number of spins $N$.

Here we must already make an important distinction, in whether one is interested in the time $T$ past which the many-body wavefunction has appreciable weight on the exact ground state (``time-to-solution'') or merely the time $T$ past which the expected energy is sufficiently close to that of the ground state (``time-to-threshold'').
Both are meaningful figures of merit for the performance of QA.
The former has been more common in the literature, but the latter has received attention as well~\cite{MunozBauza2024Scaling,Braida2024Tight} and is more amenable to our analysis.
We will thus primarily focus on the time-to-threshold in this paper, in particular how the time $T$ scales with $N$ in order for the energy to be within a fixed fraction of the ground state (i.e., for the energy \textit{density} to be within a non-zero threshold).

This scaling is closely related to whether Eq.~\eqref{eq:conventional_QA_Hamiltonian} has (ground-state) phase transitions as a function of $s$.
While exceptions are known~\cite{QuantumAdiabaticAlgorithm2012laumann,EnergyGapFirstOrder2013tsuda}, one generically finds that:
\begin{itemize}
\item If there are no phase transitions, then the time-to-threshold is $O(1)$ with respect to $N$ (regardless of the value of the threshold, although the time increases as the threshold decreases).
\item If there are only second-order phase transitions, then the time-to-threshold is polynomial in $N$.
\item If there are first-order phase transitions, then the time-to-threshold is exponential in $N$.
\end{itemize}
These relationships motivate why many studies of QA focus on the thermodynamic phase diagram for a given model --- since it is standard to classify polynomial-time algorithms as ``efficient'' and exponential-time algorithms as ``inefficient'', success is declared if the phase diagram has no worse than second-order phase transitions (and ideally none at all).
Unfortunately, it has been found that first-order phase transitions are quite common under conventional QA~\cite{SimpleGlassModels2008jorg,Young2010First,Jorg2010Energy,FirstOrderTransitionsPerformance2010jorgb,Hen2011Exponential,QuantumMeanfieldModels2012bapst,Farhi2012Performance,Altshuler2010Anderson,Foini2010Solvable,Bapst2013Quantum,Knysh2016ZeroTemperature,Baldwin2018Quantum}.

IQA attempts to circumvent phase transitions by modifying the Hamiltonian to
\begin{equation} \label{eq:inhomogeneous_QA_Hamiltonian_v1}
H(s, \tau) = s H_0(\hat{\sigma}^z) - \sum_{j=1}^{(1-\tau)N} \hat{\sigma}_j^x.
\end{equation}

There are now two independent parameters: $s$ controls the relative strength of $H_0$ compared to the fields, exactly as for conventional QA, while $\tau$ determines how many spins experience any transverse field at all.
Varying $\tau$ thus corresponds to turning off fields one by one.
The desired ground state is at $s = \tau = 1$, while the easy-to-prepare ground state is at $s = \tau = 0$.

Ref.~\cite{ExponentialSpeedupQuantum2018susa} has determined the $s$-$\tau$ phase diagram for a particular solvable choice of $H_0$ (the $p$-spin model defined below).
Although conventional QA suffers from a first-order transition in this model, IQA remarkably avoids phase transitions altogether along appropriate paths in the $s$-$\tau$ plane.
While this result was later found to not be particularly robust (say to non-zero temperature or only partially turning off fields)~\cite{QuantumAnnealing$p$spin2018susa}, it is extremely encouraging for IQA.
Given the aforementioned relationship between phase diagrams and annealing timescales, this suggests that IQA may be able to reach the ground state (to within any $O(1)$ percentage) in $O(1)$ time.

Here we show that this relationship is deceptive as applied to IQA, however.
Interestingly, IQA violates both the first and third points: it requires $O(N)$ time to approach the ground state even in the absence of transitions, and it will often have first-order transitions which render the time-to-threshold not merely exponential but \textit{infinite} (in that the ground state cannot be approached on any timescale).
While the explanation is rather straightforward, this demonstrates that when considering QA beyond the conventional approach, one cannot take for granted the connection between performance and the phase diagram.

To understand these claims, first note that while the thermodynamic-limit ($N \rightarrow \infty$) phase diagram of Eq.~\eqref{eq:inhomogeneous_QA_Hamiltonian_v1} makes sense for any $\tau \in [0, 1]$, IQA as a dynamical process will always involve a finite number of spins, and Eq.~\eqref{eq:inhomogeneous_QA_Hamiltonian_v1} is only well-defined for $\tau$ a multiple of $1/N$.
This poses a problem for how to even interpret IQA, as it does not allow for one to vary $\tau$ smoothly as a function of time.
Instead, as already noted in the original Ref.~\cite{ExponentialSpeedupQuantum2018susa}, one should continuously turn off each individual field.
We thus replace Eq.~\eqref{eq:inhomogeneous_QA_Hamiltonian_v1} with
\begin{equation} \label{eq:inhomogeneous_QA_Hamiltonian_v2}
H(s, \tau) = s H_0(\hat{\sigma}^z) - \sum_{j=1}^N \Gamma_j(\tau) \hat{\sigma}_j^x,
\end{equation}
where
\begin{equation} \label{eq:inhomogeneous_QA_field_schedule}
\Gamma_j(\tau) = \begin{cases} 1, \quad &0 \leq \tau < \frac{N-j}{N} \\ 1 - N \big( \tau - \frac{N-j}{N} \big), \quad &\frac{N-j}{N} \leq \tau < \frac{N-j+1}{N} \\ 0, \quad &\frac{N-j+1}{N} \leq \tau \leq 1 \end{cases}.
\end{equation}
In words, $\Gamma_j(\tau)$ interpolates from 1 to 0 as $\tau$ increases from $(N-j)/N$ to $(N-j+1)/N$.
Again, this distinction is unnecessary for the phase diagram, since Eqs.~\eqref{eq:inhomogeneous_QA_Hamiltonian_v2} \&~\eqref{eq:inhomogeneous_QA_field_schedule} become equivalent to Eq.~\eqref{eq:inhomogeneous_QA_Hamiltonian_v1} as $N \rightarrow \infty$, but it is essential when considering the dynamics.
Furthermore, it suggests that the timescale needed to remain near the ground state might be dictated by each \textit{individual} field turning off sufficiently slowly.
Since the fields are turned off one by one, this translates to an additional factor of $N$ in the total time $T$.
We explicitly confirm this to be the case in what follows.

Unfortunately, the situation is likely to be even worse in hard non-mean-field problems (at least those which are diagonal in the $z$ basis).
This stems from the observation that once the field on spin $j$ is turned off, $\hat{\sigma}_j^z$ becomes a conserved quantity and its expectation value will no longer change in time.
Thus spins can very well ``freeze'' in the wrong direction, especially those whose fields are turned off near the beginning of the protocol.
In terms of the level structure, this translates to exact level crossings at which the gap becomes identically zero, and thus the system cannot remain in the ground state regardless of $T$.
This is reflected both in the statement of the adiabatic theorem and simply by visualizing the Hamiltonian as block-diagonal with respect to $\hat{\sigma}_j^z$ --- clearly there can be no transitions from one block to the other if there are no matrix elements connecting the two.

We provide evidence for these two main results --- that IQA requires $O(N)$ time even when there are no phase transitions, and that it is likely to fail regardless of the timescale in hard problems --- in Secs.~\ref{sec:mean_field_theory} and~\ref{sec:finite_size_numerics} respectively.
The former relies on large-$N$ dynamical mean-field calculations applied to the same model as in Refs.~\cite{ExponentialSpeedupQuantum2018susa,QuantumAnnealing$p$spin2018susa}, and the latter turns to exact diagonalization of small-$N$ spin-glass problems.
Beforehand, in Sec.~\ref{sec:general_argument}, we simply formalize the observation made above --- assuming that $H_0$ is diagonal in the $z$ basis, the expectation value of $\hat{\sigma}_j^z$ cannot change once field $j$ is turned off.
We lastly close with directions for future work in Sec.~\ref{sec:conclusion}.

\section{General argument} \label{sec:general_argument}

Consider the IQA Hamiltonian given by Eqs.~\eqref{eq:inhomogeneous_QA_Hamiltonian_v2} \&~\eqref{eq:inhomogeneous_QA_field_schedule}, with $s$ and $\tau$ being functions of time $t$.
To reiterate, we are assuming that $H_0$ is diagonal in the $z$ basis, as is often the case in many applications of QA.
Then for any $j \geq (1 - \tau)N + 1$, $\Gamma_j(\tau) = 0$ and so $\hat{\sigma}_j^z$ commutes with the full Hamiltonian.
With $| \Psi(t) \rangle$ denoting the state of the spins, and given that $\partial_t | \Psi(t) \rangle = -i H(s, \tau) | \Psi(t) \rangle$, we have
\begin{equation} \label{eq:spin_conservation}
\frac{\textrm{d} \langle \Psi(t) | \hat{\sigma}_j^z | \Psi(t) \rangle}{\textrm{d}t} = i \big< \Psi(t) \big| \big[ H(s, \tau), \hat{\sigma}_j^z \big] \big| \Psi(t) \big> = 0.
\end{equation}

Although simple and brief, this observation has important implications for IQA.
If the final state of the system is to be the ground state of $H_0$, then since the expectation value of $\hat{\sigma}_j^z$ is constant once $\Gamma_j = 0$, spin $j$ must be pointing in the ``correct'' direction (that of the ground state of $H_0$) at the moment when its field first turns off.
This must be true even for the first spins to lose their fields at the beginning of the protocol, when all others still experience strong fields and the term $s H_0$ is weak.

First and foremost, this suggests that IQA is likely to fail in hard problems, where frustration and disorder will presumably make it very difficult to identify which is the correct direction for the first few spins.
All others will still be polarized in the $x$ direction at that point, and thus the energy landscape for the first few spins may look very different than how it would in the ground state of $H_0$.
Note that this is unrelated to the speed at which the fields are reduced --- even if the fields are turned off sufficiently slowly and each spin settles adiabatically into a $\hat{\sigma}^z$ eigenstate, it may simply be the wrong eigenstate.

Second, even if it is possible for every spin to identify the correct direction during the period when its field is reduced, this suggests that the total time $T$ for the IQA protocol cannot be less than $O(N)$.
Having $T$ be parametrically less than $O(N)$ would correspond to each individual field being turned off rapidly in the large-$N$ limit, and thus each spin would freeze in whatever direction it happened to point at the moment when its field was suddenly quenched --- there is no reason to believe that this would even be a $\hat{\sigma}^z$ eigenstate to begin with.

Of course, these arguments (other than Eq.~\eqref{eq:spin_conservation} itself) are somewhat vague and still need to be confirmed in explicit examples.
In Sec.~\ref{sec:mean_field_theory}, we study the dynamics of the ferromagnetic $p$-spin model using mean-field techniques (technical details can be found in the appendices), and confirm that the time-to-threshold is specifically $O(N)$.
In Sec.~\ref{sec:finite_size_numerics}, we then study small disordered problems (namely the Sherrington-Kirkpatrick model~\cite{Fischer1991,Mezard2009}) via exact diagonalization.
We show that IQA can indeed fail due to spins freezing in the wrong direction, and that this manifests as genuine (not merely avoided) level crossings in the spectrum.
We find that the fraction of problem instances with at least one level crossing increases rapidly to 1 as the system size increases.

\section{Mean-field theory} \label{sec:mean_field_theory}

\begin{figure}[t]
\centering
\includegraphics[width=0.47\textwidth]{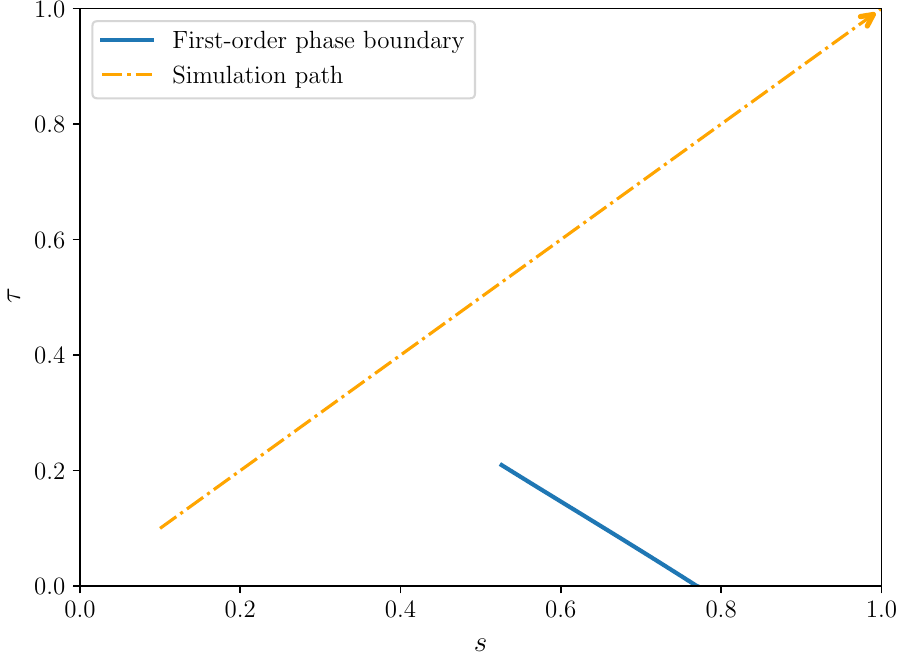}
\caption{Phase diagram of the $p$-spin model ($p = 3$) under IQA (see Ref.~\cite{ExponentialSpeedupQuantum2018susa}). The solid blue line indicates a first-order phase transition. Note that there are paths from $s = \tau = 0$ (where $H(s, \tau) = -\sum_{j=1}^N \hat{\sigma}_j^x$) to $s = \tau = 1$ (where $H(s, \tau) = H_0$) which avoid transitions altogether. We specifically study the dynamics of IQA along the path indicated by the dashed orange line~\cite{starting_point_note}.}
\label{fig:fig0}
\end{figure}

In this section, we analyze the performance of IQA in the ferromagnetic $p$-spin model, for which
\begin{equation} \label{eq:p_spin_Hamiltonian}
H_0 = -N \Bigg( \frac{1}{N} \sum_{j=1}^{N} \hat{\sigma}_j^z \Bigg)^p.
\end{equation}
Equivalently, the model consists of equal-strength interactions among all sets of $p$ spins (the factors of $N$ are to ensure that each spin experiences a net interaction of order 1 as $N$ increases).
Even though the ground state of this model is trivial (all spins point in the $z$ direction), it is a common and useful test case for QA protocols, as it has a non-trivial phase diagram in the presence of transverse fields while still being exactly solvable in the large-$N$ limit~\cite{Jorg2010Energy,QuantumMeanfieldModels2012bapst}.
For concreteness, we set $p = 3$ throughout (other values of $p$ behave similarly).
The thermodynamics of IQA for this model, meaning the ground-state phase diagram of Eq.~\eqref{eq:inhomogeneous_QA_Hamiltonian_v1}, was determined in Refs.~\cite{ExponentialSpeedupQuantum2018susa,QuantumAnnealing$p$spin2018susa}.
The main result is as shown in Fig.~\ref{fig:fig0}.
Note that there are indeed paths in the $s$-$\tau$ plane from $(0, 0)$ to $(1, 1)$ which avoid phase transitions.
As discussed in Sec.~\ref{sec:introduction}, this suggests that one can begin with all spins pointing in the $x$ direction ($s = \tau = 0$) and adiabatically evolve them into the ground state of $H_0$ ($s = \tau = 1$) in a reasonably short time.

Here we directly study the dynamics of IQA: we solve the Schrodinger equation
\begin{equation} \label{eq:Schrodinger_equation}
\partial_t \big| \Psi(t) \big> = -i H \big( s(t), \tau(t) \big) \big| \Psi(t) \big>,
\end{equation}
where $s(t)$ and $\tau(t)$ follow the path indicated in Fig.~\ref{fig:fig0} over time $T$, namely~\cite{starting_point_note}
\begin{equation} \label{eq:p_spin_specific_parameter_schedule}
s(t) = \frac{1}{10} + \frac{9t}{10T}, \qquad \tau(t) = \frac{1}{10} + \frac{9t}{10T}.
\end{equation}
The initial state is the ground state of $H(s(0), \tau(0))$, and we determine how large $T$ must be in order for the energy of the final state to be sufficiently close to the ground state of $H_0$.

Much as the thermodynamics does, the dynamics simplifies greatly at large $N$.
While the derivation is somewhat formal (see App.~\ref{sec:appendix2}), the end result is straightforward to state.
The system remains in a product state throughout the protocol, i.e., $| \Psi(t) \rangle = \bigotimes_j | \psi_j(t) \rangle$, and each individual spin evolves under a ``mean field'' $h(t)$ in addition to whatever transverse field it experiences:
\begin{equation} \label{eq:mean_field_Schrodinger_equation}
\begin{aligned}
\partial_t \big| \psi_j(t) \big> &= i \Big[ s(t) h(t) \hat{\sigma}_j^z + \Gamma_j \big( \tau(t) \big) \hat{\sigma}_j^x \Big] \big| \psi_j(t) \big> \\
&\equiv -i H_{\textrm{eff},j}(t) \big| \psi_j(t) \big>,
\end{aligned}
\end{equation}
where
\begin{equation} \label{eq:p_spin_mean_field_expression}
h(t) = p m^z(t)^{p-1},
\end{equation}
with magnetization
\begin{equation} \label{eq:average_magnetization_definition}
m^z(t) \equiv \frac{1}{N} \sum_{j=1}^N \big< \psi_j(t) \big| \hat{\sigma}_j^z \big| \psi_j(t) \big>.
\end{equation}
The initial state $| \psi_j(0) \rangle$ is obtained from the thermodynamic calculation (see App.~\ref{sec:appendix1}).
We determine the magnetization $m^z(0)$ via Eq.~\eqref{eq:average_magnetization_definition}, then $h(0)$ via Eq.~\eqref{eq:p_spin_mean_field_expression}, and lastly use that value of $h(0)$ to compute $| \psi_j(\Delta t) \rangle \approx \exp{[-i H_{\textrm{eff},j}(0) \Delta t]} | \psi_j(0) \rangle$ (for all $j$ and using sufficiently small $\Delta t$) according to Eq.~\eqref{eq:mean_field_Schrodinger_equation}.
We repeat this process to determine $| \psi_j(2 \Delta t) \rangle$, then $| \psi_j(3 \Delta t) \rangle$, and so on.

The expectation value of $H_0$ is given simply by
\begin{equation} \label{eq:p_spin_mean_field_energy}
\frac{\big< \Psi(t) \big| H_0 \big| \Psi(t) \big>}{N} = -m^z(t)^p.
\end{equation}
Thus $m^z(t)$ serves as a useful proxy for how close $| \Psi(t) \rangle$ is to the ground state of $H_0$.
In particular, the expected energy is within a certain threshold of the ground state if and only if the magnetization is within a corresponding threshold of 1 (this is why it is more natural for us to consider the time-to-threshold in this work).

Technically these ``mean-field equations'' are exact only for an infinite number of spins.
Nonetheless, we use them at large but finite $N$ so as to understand how the dynamical behavior depends on $N$.
As discussed in Sec.~\ref{sec:introduction}, we specifically focus on the time $T$ required for the magnetization at the end of the IQA protocol to be within a certain threshold of 1, and how that value of $T$ scales with $N$.

\begin{figure}[t]
\centering
\includegraphics[width=0.47\textwidth]{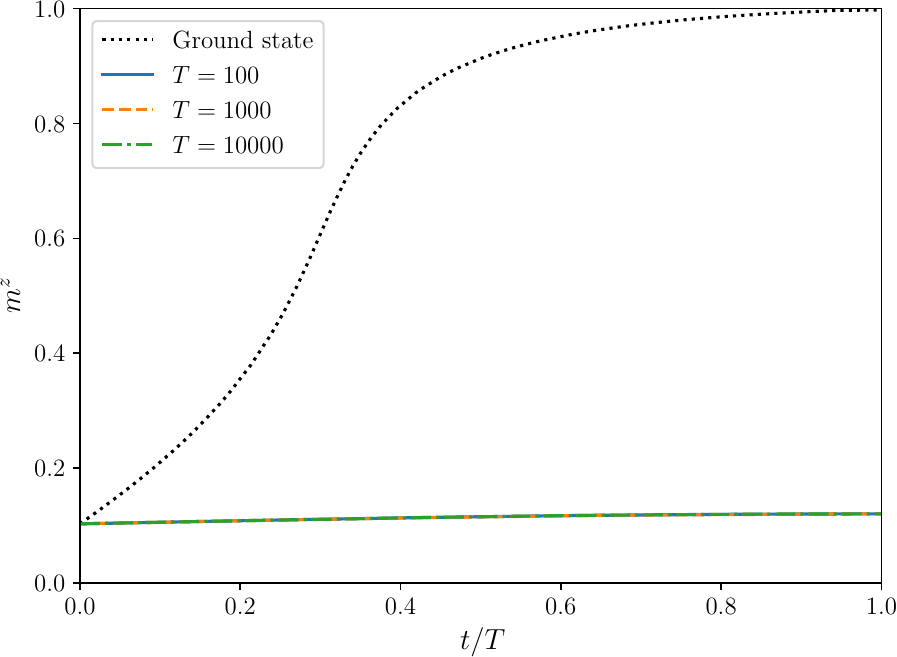}
\caption{Magnetization as a function of time during the sudden-quench protocol of Eq.~\eqref{eq:inhomogeneous_QA_quench_schedule}. We use $N = 5000$ and a time-step of $\Delta t = 0.01$. Different curves correspond to different total times $T$ as indicated, and the dotted line gives the magnetization in the ground state of $H(s(t), \tau(t))$ for comparison. The magnetization for different values of $T$ differs on the order of $10^{-5}$, imperceptible on this scale.}
\label{fig:fig1}
\end{figure}

We give results for three particular cases.
The first technically uses a different form for the transverse fields than Eq.~\eqref{eq:inhomogeneous_QA_field_schedule}: rather than have $\Gamma_j(\tau)$ linearly interpolate from 1 to 0, we turn it off suddenly at $\tau = (N-j)/N$, i.e.,
\begin{equation} \label{eq:inhomogeneous_QA_quench_schedule}
\Gamma_j(\tau) = \begin{cases} 1, \quad &0 \leq \tau < \frac{N-j}{N} \\ 0, \quad &\frac{N-j}{N} \leq \tau \leq 1 \end{cases}.
\end{equation}
While it will be a useful reference point, this ``sudden-quench'' protocol is unlikely to work, and we confirm in Fig.~\ref{fig:fig1} that it indeed does not for the $p$-spin model.
Plotted is the magnetization $m^z$ as a function of time $t$, for various total times $T$ (the magnetization of the instantaneous ground state of $H(s(t), \tau(t))$ is also shown for comparison).
As discussed above, if this protocol were able to succeed in locating the desired ground state, then $m^z(T)$ would approach 1 as $T$ increases.
Clearly it does not.
In fact, the magnetization is extremely insensitive to the total time $T$ --- all three curves are indistinguishable on the scale shown.

\begin{figure}[t]
\centering
\includegraphics[width=0.47\textwidth]{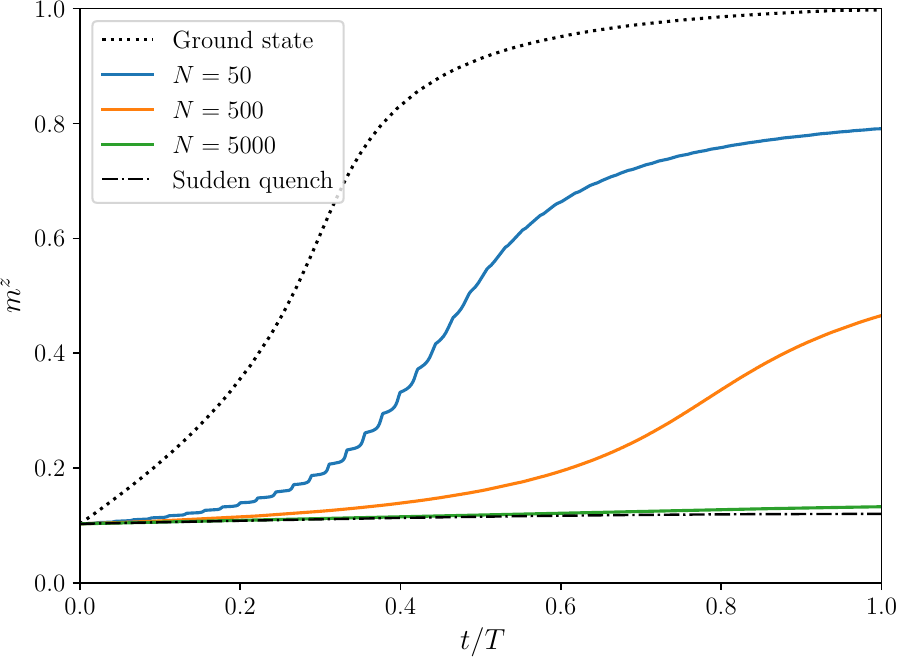}
\vskip 12pt
\includegraphics[width=0.47\textwidth]{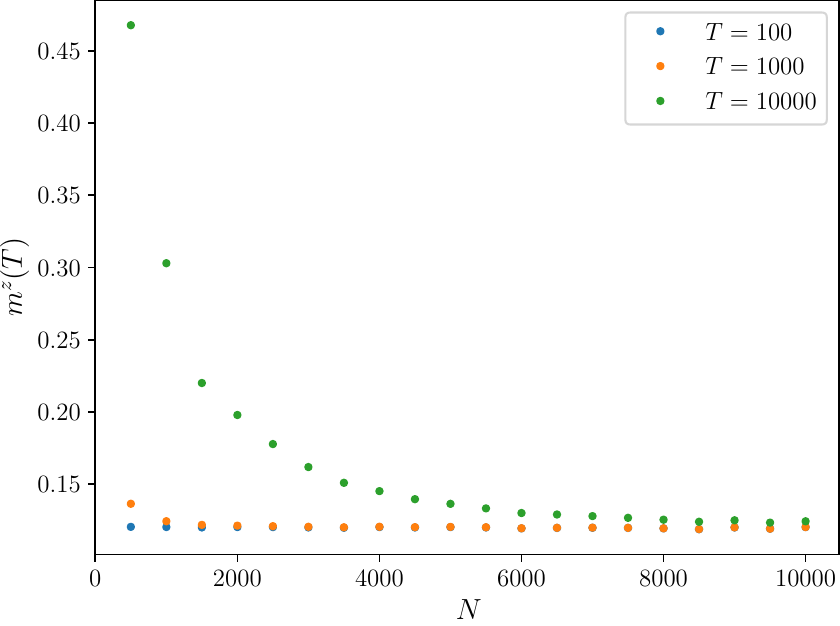}
\caption{(Top) Magnetization as a function of time during IQA (with fields given by Eq.~\eqref{eq:inhomogeneous_QA_field_schedule}), for various $N$ at fixed $T = 10000$. The dotted line gives the magnetization in the instantaneous ground state, and the dash-dotted line gives the magnetization during the sudden-quench protocol from Fig.~\ref{fig:fig1}. The time-step is $\Delta t = 0.1$. (Bottom) Final magnetization $m^z(T)$ as a function of $N$ for various $T$.}
\label{fig:fig2}
\end{figure}

Next return to the more sensible IQA protocol, where the fields are given by Eq.~\eqref{eq:inhomogeneous_QA_field_schedule}, but with $T$ which is $O(1)$ with respect to $N$.
This is delicate, since for any \textit{fixed} $N$, the magnetization will always approach that of the ground state for sufficiently large $T$.
Instead we must increase $N$ at fixed $T$, extract the asymptotic large-$N$ value of the magnetization, and then study how that asymptotic value changes as $T$ increases.
With this order of limits, one might expect that the protocol would behave similarly to the sudden quench of Fig.~\ref{fig:fig1}.
The top panel of Fig.~\ref{fig:fig2} shows a representative example confirming that this is the case.
The magnetization is plotted versus time for various $N$ at a specific fixed value of $T$, and while the curve roughly resembles that of the ground state at smaller $N$, it unambiguously approaches the sudden-quench curve as $N$ increases.

The bottom panel of Fig.~\ref{fig:fig2} gives the final magnetization $m^z(T)$ as a function of both $N$ and $T$.
Regardless of $T$, the final magnetization decreases towards an asymptotic value quite far from 1 as $N$ increases.
That value is roughly independent of $T$ as well, consistent with the claim that the continuous IQA protocol behaves like the sudden quench at large $N$.
Thus in short, IQA fails to approach the ground state on any $O(1)$ timescale.

\begin{figure}[t]
\centering
\includegraphics[width=0.47\textwidth]{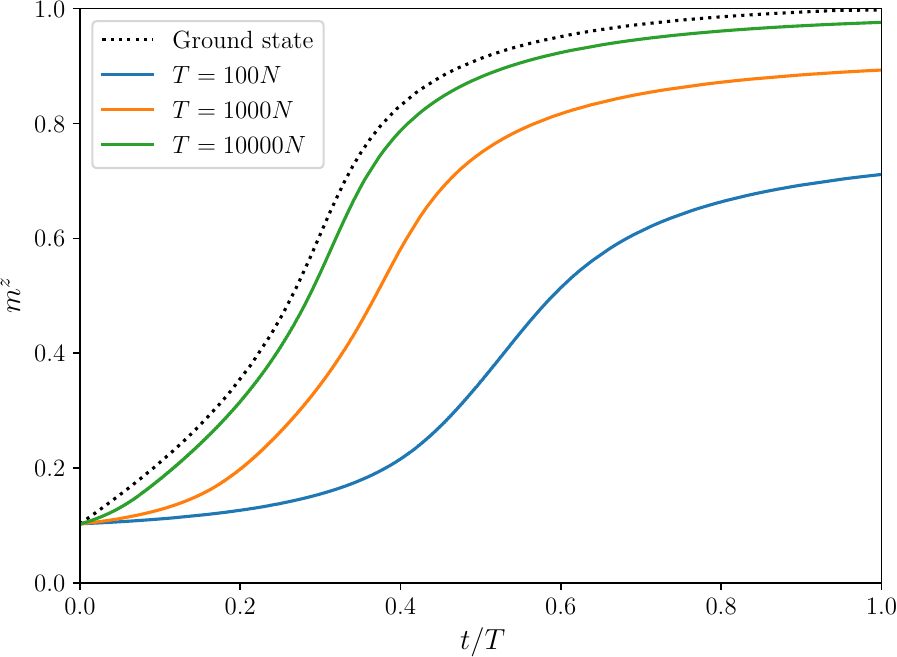}
\vskip 12pt
\includegraphics[width=0.47\textwidth]{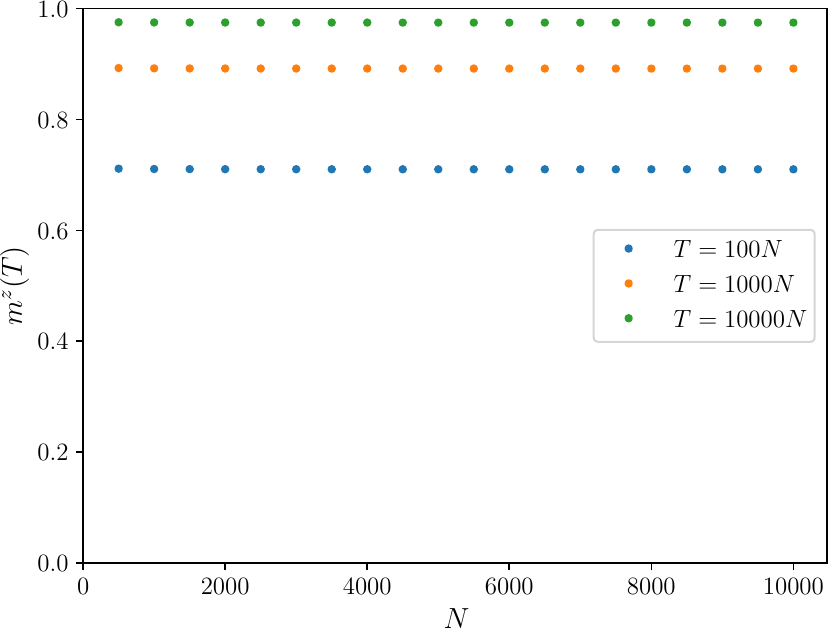}
\caption{(Top) Magnetization as a function of time during IQA, for various $T$ at fixed $N = 500$. The dotted line gives the magnetization in the instantaneous ground state for comparison. The time-step is $\Delta t = 0.1$. (Bottom) Final magnetization $m^z(T)$ as a function of $N$, for various $T$ which are themselves proportional to $N$.}
\label{fig:fig3}
\end{figure}

Lastly consider $T = O(N)$.
More specifically, we set $T = cN$ for some fixed $c$, increase $N$ and extract the asymptotic value of $m^z(T)$, and then study how that value changes as $c$ increases.
The results are shown in the bottom panel of Fig.~\ref{fig:fig3} (the top panel gives examples of magnetization curves as in Fig.~\ref{fig:fig2}).
Interestingly, $m^z(T)$ is largely independent of $N$ under this scaling, with a value that indeed approaches 1 as $c$ increases.
This confirms that $T = O(N)$ is sufficient for IQA to succeed in approaching the ground state, and in fact, it confirms that $T = O(N)$ is \textit{necessary} as well --- were it possible to reach the ground state in parametrically smaller $T$, then $m^z(cN)$ would instead flow towards 1 as $N$ increases.

Thus despite following a path in the phase diagram that avoids transitions altogether (see Fig.~\ref{fig:fig0}), IQA requires $O(N)$ time even merely to reach energies within a finite fraction of the ground state.
This stands in contrast to conventional QA, which (at least when there are no phase transitions) requires only $O(1)$ time.
To be sure, the additional factor of $N$ pales in comparison to the exponential time that would be needed to cross a first-order transition, so one can still consider IQA applied to the $p$-spin model as a success.
Nonetheless, for large problems, this does constitute a substantial slowdown that should be kept in mind.

Before proceeding, it is interesting to note that the time-to-solution for IQA does \textit{not} suffer from the same factor of $N$, at least not according to the upper bounds established by the adiabatic theorem.
This is not because the time-to-solution is any shorter for IQA, but rather because the time-to-solution for conventional QA already contains factors of $N$.
To be concrete, consider varying $s$ and $\tau$ along the diagonal line $s(t) = \tau(t)$.
The heuristic statement of the adiabatic theorem~\cite{AdiabaticQuantumComputation2018albash} claims that the system will remain in the ground state throughout the protocol if $T \gg \max_s \lVert \partial_s H(s, s) \rVert / \Delta(s)^2$, where $\Delta(s)$ is the gap of $H(s, s)$ and $\lVert \, \cdot \, \rVert$ denotes the operator norm (largest eigenvalue in absolute value).
Using Eq.~\eqref{eq:inhomogeneous_QA_Hamiltonian_v2} for $H(s, s)$ and bounding $\lVert \partial_s H(s, s) \rVert$ by the triangle inequality, we can say that the system remains adiabatic if
\begin{equation} \label{eq:adiabatic_condition}
T \gg \max_s \frac{1}{\Delta(s)^2} \bigg( \big\lVert H_0 \big\rVert + \sum_{j=1}^N \Big| \frac{\textrm{d} \Gamma_j(s)}{\textrm{d}s} \Big| \bigg).
\end{equation}
Assume that the protocol does not cross any phase transitions, so that $\Delta(s) = O(1)$.
Conventional QA has $\Gamma_j(s) = 1-s$, and thus the right-hand side of Eq.~\eqref{eq:adiabatic_condition} is $O(N)$ (note that $\lVert H_0 \rVert = N$).
IQA has $\Gamma_j(s)$ given by Eq.~\eqref{eq:inhomogeneous_QA_field_schedule}, and so $|\textrm{d}\Gamma_j(s)/\textrm{d}s| = N \delta_{j, j_0(s)}$, where $j_0(s)$ is the value of $j$ for which $s \in [(N-j)/N, (N-j+1)/N)$.
Thus the right-hand side of Eq.~\eqref{eq:adiabatic_condition} is still $O(N)$, hinting that the time-to-solution may be comparable for conventional QA and IQA.
A similar conclusion holds using rigorous formulations of the adiabatic theorem~\cite{AdiabaticQuantumComputation2018albash}, as they still involve the norms of derivatives of $H(s, s)$.
Keep in mind that this is only an upper bound on the time-to-solution, and so it does not constitute a proof, but it is certainly suggestive and warrants further investigation.

\section{Non-mean-field problems} 
\label{sec:finite_size_numerics}

We now consider non-mean-field systems with more complicated Hamiltonians, and show that IQA is likely to fail altogether when applied to such models.
Interestingly, this failure is even more problematic than that of conventional QA: there is no timescale on which IQA can reach the ground state, even exponential.
As a result, we can demonstrate the failure on small systems accessible via exact diagonalization, and no extrapolation to larger $N$ is required.

\begin{figure}[t]
\centering
\includegraphics[width=0.47\textwidth]{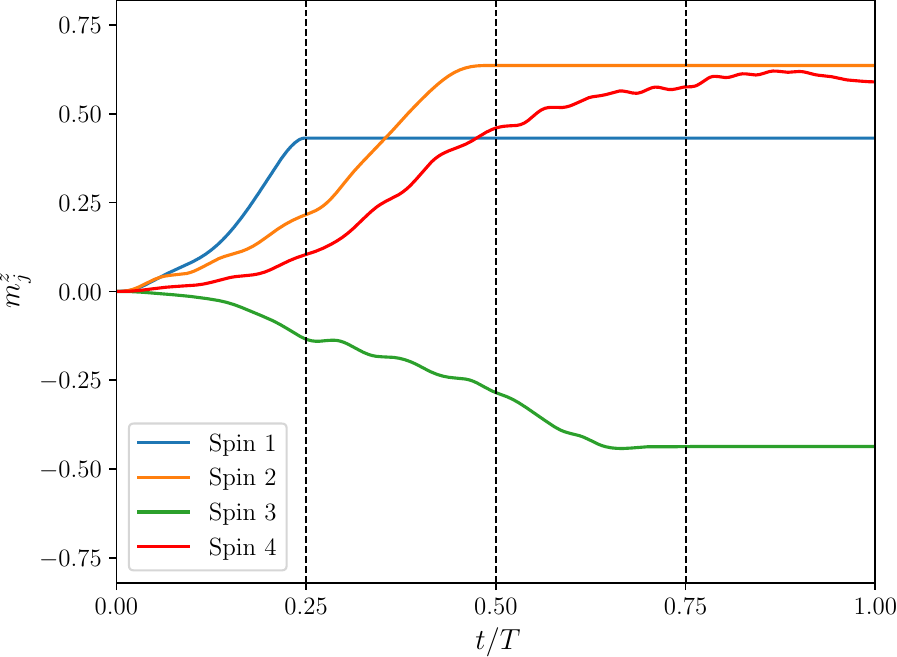}
\caption{Individual magnetizations $m_j^z \equiv \langle \hat{\sigma}_j^z \rangle$ during the IQA protocol, for $N = 4$ and $T = 10$. Vertical dashed lines indicate the times at which a transverse field is fully turned off. Time-step is $\Delta t = 0.1$ for numerical integration of the Schrodinger equation. The problem Hamiltonian $H_0$ is given by Eq.~\eqref{eq:SK_Hamiltonian} with $J_{jk} = \cos(j^4) + \cos(k^4)$ and $h_j= \cos(j^2)$.}
\label{fig:fig4}
\end{figure}

We specifically take $H_0$ to be of the form
\begin{equation} \label{eq:SK_Hamiltonian}
H_0 = - \sum_{j=1}^{N} \sum_{k=j+1}^N J_{jk} \hat{\sigma}_j^z \hat{\sigma}_k^z - \sum_{j=1}^N h_j \hat{\sigma}_j^z.
\end{equation}
Depending on the context, we either use fixed but relatively structureless expressions for $J_{jk}$ and $h_j$ (to aid in reproducibility) or draw each coefficient randomly from an independent Gaussian distribution (this latter case goes by the name of the Sherrington-Kirkpatrick model in the spin-glass literature~\cite{Fischer1991,Mezard2009}).
The full Hamiltonian is again given by Eqs.~\eqref{eq:inhomogeneous_QA_Hamiltonian_v2} and~\eqref{eq:inhomogeneous_QA_field_schedule}, now simply with $s(t) = \tau(t) = t/T$.

To begin, Fig.~\ref{fig:fig4} gives an example with $N = 4$ (obtained by directly integrating the Schrodinger equation) illustrating the result proven in Sec.~\ref{sec:general_argument}: once field $j$ is turned off, the magnetization of spin $j$, $m_j^z(t) \equiv \langle \Psi(t) | \hat{\sigma}_j^z | \Psi(t) \rangle$, becomes constant in time.
This poses a clear problem for IQA, as there is no mechanism for $m_j^z$ to ``correct itself'' in response to changes in the other spins.

\begin{figure}[t]
\centering
\includegraphics[width=0.47\textwidth]{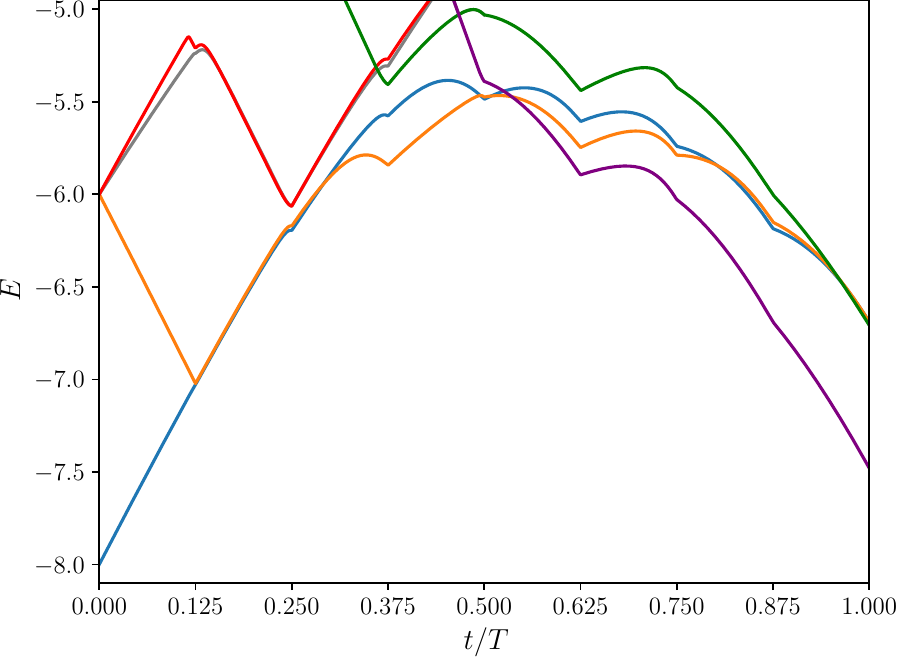}
\caption{Lowest portion of the instantaneous spectrum of $H(s(t), \tau(t))$ during the IQA protocol, for $H_0$ given by Eq.~\eqref{eq:SK_Hamiltonian} with $J_{jk} = \cos(j^5 + k^5)/2$ and $h_j = \cos(j)$ ($N = 8$). Curves are colored to aid in visualizing level crossings.}
\label{fig:fig5}
\end{figure}

The problem becomes even more apparent when viewed through the level structure.
An example is shown in Fig.~\ref{fig:fig5}, which plots the lowest few eigenvalues of $H(s(t), \tau(t))$ throughout the IQA protocol (treating $t/T$ as a parameter and using exact diagonalization).
The curves are colored simply to aid in visualization.
We see that there are many \textit{exact} level crossings at which the gap becomes identically zero, even for this small system size ($N = 8$).
While it may at first be surprising to see a finite-size gap which is exactly zero --- studies of QA rather tend to find avoided level crossings in which the gap merely decreases to zero as $N \rightarrow \infty$ --- note that there is one situation where this naturally occurs: Hamiltonians which are block-diagonal, such that a level in one block crosses a level in another.
This is precisely what happens in IQA: once $\Gamma_j(\tau) = 0$, $H(s, \tau)$ commutes with $\hat{\sigma}_j^z$ and thus is block-diagonal with respect to the eigenspaces of the latter.
For each of the level crossings indicated in Fig.~\ref{fig:fig5}, we have confirmed that the two states involved in the crossing indeed belong to different blocks, in that they differ in the eigenvalue of at least one frozen $\hat{\sigma}_j^z$ (this also gives a way to unambiguously identify the crossing as exact).

The adiabatic theorem does not apply if the gap becomes identically zero, and since the levels which cross are associated with different blocks of the Hamiltonian, it is clear that the state of the system will (in the limit $T \rightarrow \infty$) simply follow a level through the crossing.
Thus, for the example in Fig.~\ref{fig:fig5}, IQA in the limit $T \rightarrow \infty$ will follow the blue curve and thus end up in the second excited state (although it is hard to distinguish in the figure).

\begin{figure}[t]
\centering
\includegraphics[width=0.47\textwidth]{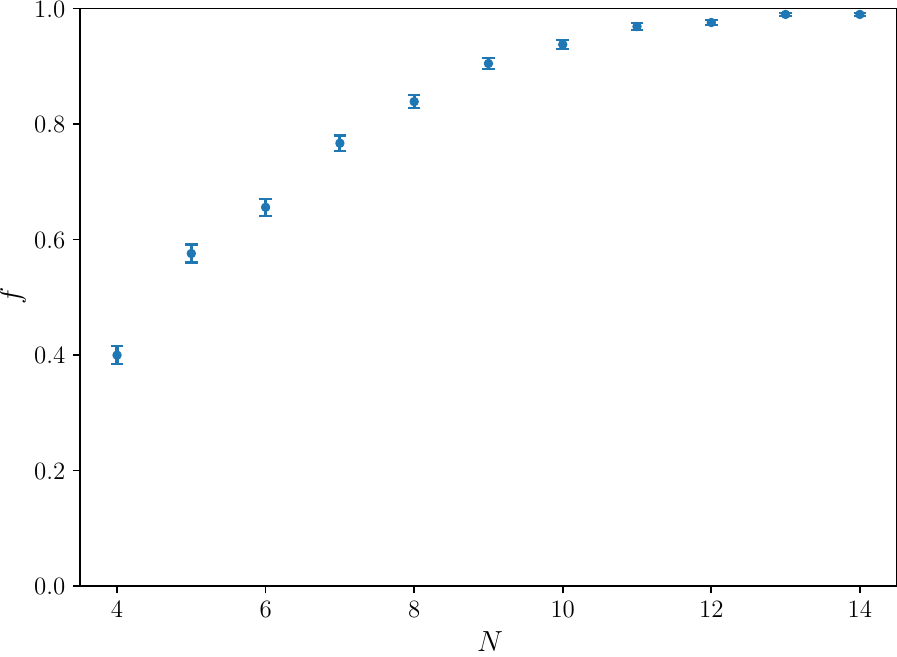}
\caption{Fraction $f$ of random realizations of Eq.~\eqref{eq:SK_Hamiltonian} (each $J_{jk}$ and $h_j$ drawn independently from a Gaussian with mean zero and variance $1/N$) for which there is at least one level crossing during the IQA protocol. 1000 realizations were used for each value of $N$.}
\label{fig:fig6}
\end{figure}

The remaining question is how common these level crossings are in typical problem instances.
For each $N$ ranging from 4 to 14, we have generated 1000 random instances of $H_0$ by drawing each $J_{jk}$ and $h_j$ from a Gaussian (mean zero and variance $1/N$), and counted the fraction that have at least one level crossing during the IQA protocol.
The results are shown in Fig.~\ref{fig:fig6}.
The fraction $f$ is actually somewhat low for the smallest sizes considered, but it rapidly increases with $N$ and exceeds $99\%$ by $N = 14$.
While one should consider whether a more realistic problem might have additional structure which is not captured by these random instances, this ensemble (known as the Sherrington-Kirkpatrick model) is a standard prototype for many spin-glass and optimization problems~\cite{Fischer1991,Mezard2009}.
We thus find it extremely unlikely that a typical problem instance at large $N$ would not have any level crossing, and thus unlikely that IQA would succeed in finding the ground state (on any timescale).

\begin{figure}[t]
\centering
\includegraphics[width=0.47\textwidth]{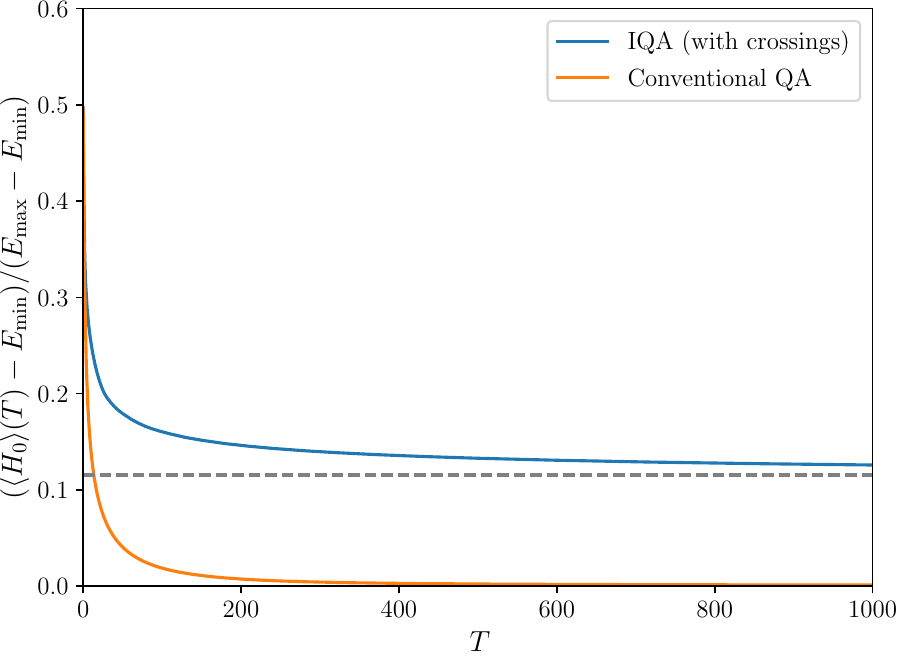}
\caption{Energy at the end of the annealing protocol (IQA in blue and conventional QA in orange) as a function of the total protocol time $T$, averaged over random realizations of Eq.~\eqref{eq:SK_Hamiltonian} (each $J_{jk}$ and $h_j$ drawn independently from a Gaussian with mean zero and variance $1/N$). The energy is given as a fraction of the spectral width, i.e., the $y$-axis is $(\langle H_0 \rangle(T) - E_{\textrm{min}})/(E_{\textrm{max}} - E_{\textrm{min}})$, where $E_{\textrm{min}}$ is the lowest eigenvalue of $H_0$, $E_{\textrm{max}}$ is the highest eigenvalue, and the expectation value of $H_0$ is taken in the state at the end of the protocol. The dashed line indicates the value of the final energy under IQA for $T = 10000$, which we take as an estimate of the asymptotic large-$T$ value (in going from $T = 5000$ to $10000$, the energy changes only in the third decimal place). For the averaging over realizations, 1000 realizations were initially drawn, but only those for which the ground state exhibits at least one level crossing under IQA have been included in the averaging. System size is $N = 8$, and the time-step for numerical integration of the Schrodinger equation is $\Delta t = 0.01$.}
\label{fig:fig7}
\end{figure}

% The average final energy measured as a fraction of the spectral width away from the ground state after each realization evolves for a time $T$ using the IQA protocol. Specifically, for each realization, we compute the expectation value $\langle\hat{H}_0\rangle$ of the final state, subtract the ground state energy of the problem Hamiltonian $E_0$, and divide by the spectral width $(E_{\text{max}} - E_0)$, where $E_{\text{max}}$ is the highest excited state energy. The result, $(\langle \varphi_{(t=T)} | H_0 | \varphi_{(t=T)} \rangle - E_0)/(E_{\text{max}} - E_0)$, is then averaged over all realizations. These values are computed at time intervals in steps of 1 from $T = 0$ to $T = 100$, and in steps of 100 from $T = 100$ to $T = 5000$. For comparison, the corresponding values obtained from conventional QA are also shown. Time-step is $\Delta t = 0.01$ for numerical integration of the Schrodinger equation.

While technically the discussion in this section has been for the time-to-\textit{solution}, it is clear that a similar conclusion applies to the time-to-threshold: since the system will end at an excited energy even in the $T \rightarrow \infty$ limit, the time-to-threshold is infinite for any energy below that excited value. To emphasize this point, and to draw an explicit comparison to conventional QA, we plot in Fig.~\ref{fig:fig7} the expectation value of $H_0$ at the end of the annealing protocol (both IQA and conventional QA) as a function of the total time $T$, using the same random realizations of $H_0$ as for Fig.~\ref{fig:fig6} (at $N = 8$). The result is given as a fraction of the spectral width (difference between largest and smallest eigenvalues of $H_0$), and it is averaged over those realizations that have at least one level crossing~\cite{crossing_average_note}. Although the final energy under conventional QA approaches zero at large $T$, the energy under IQA approaches a value at roughly 12\% of the spectral width (as compared to the initial energy at roughly 50\%). The fact that this final energy is appreciably different from both 0\% and 50\% is not surprising --- the first spins to have their fields turned off freeze in essentially random directions, since the others at that point remain polarized largely in the $x$ direction, but the last spins freeze into their lowest-energy configurations (conditioned on the random directions of the first spins).

\section{Conclusion} \label{sec:conclusion}

We have assessed the performance of inhomogeneous quantum annealing (IQA), primarily the time required for the final energy to be within a finite fraction of the ground state (``time-to-threshold''), and found that it is anomalously slow compared to what the phase diagram would suggest.
Interestingly, this happens both in the presence and absence of phase transitions.
Even when there are no transitions, the time-to-threshold for IQA is $O(N)$ (whereas it would be $O(1)$ for conventional QA in the absence of transitions).
Moreover, first-order transitions are likely to be quite common in hard problems, and they are especially problematic because they prevent IQA from reaching low-energy states on any timescale (rather than merely exponential as for conventional QA).
We explain both of these results through the simple observation that once an individual field is turned off during the IQA protocol, the $z$ magnetization of the corresponding spin becomes frozen.
Spins can very well freeze in the wrong direction, which gives rise to phase transitions at which the gap is identically zero, and even if this does not occur, each individual field must be turned off slowly in order for the corresponding spin to adiabatically settle into a $\hat{\sigma}^z$ eigenstate before freezing (thus the total runtime can be no better than $O(N)$).

Despite these negative results, particularly the prevalence of first-order transitions with strictly vanishing gap, there are a number of reasons to investigate IQA further rather than abandon it.
First and foremost, we have only considered IQA under rather idealized operating conditions, in which the system evolves under perfectly isolated dynamics without any noise.
This is certainly not the case for present-day annealing platforms.
Since the major issue with ideal IQA is spins freezing in the wrong direction, it is conceivable that noise could improve the performance of IQA by causing spin-flips out of incorrect states, and thus experimental annealing machines might actually perform \textit{better} than our results suggest.
Experimental implementations of protocols closely related to (although not identical to) IQA have indeed observed significant improvements over conventional QA~\cite{ExperimentalDemonstrationPerturbative2017lanting,InhomogeneousDrivingQuantum2020adame}.
Even as a purely theoretical question, it would be quite interesting to investigate the dynamics of IQA under noise and other imperfections, to see if and how they resolve the problem of frozen spins (see Ref.~\cite{QuantumAnnealing$p$spin2018susa} in this context). 

Keep in mind that since the main application of QA is for optimization problems, the goal does not have to be finding the exact ground state but rather merely a lower-energy state than other algorithms could yield.
Thus even if IQA cannot reach the exact ground state due to first-order transitions, the question remains as to how low in energy it can reach.
We suspect --- although this should be checked --- that IQA will yield states which lie at non-zero energy density above the ground state, i.e., at a finite fraction of the spectral width even in the $N \rightarrow \infty$ limit.
Since other algorithms also tend to become stuck at non-zero energy densities in hard problems~\cite{Mezard2009}, there is potential for IQA to outperform competitors, at least in certain cases.
This possibility should be investigated further (note that the example in Fig.~\ref{fig:fig7} does not quite answer this since it does not address the scaling with $N$).

Lastly, the idea of approximate optimization via IQA raises a number of follow-up questions.
For example, does the order in which one turns off fields affect the final energy density, and is it possible to determine a sufficiently good order efficiently?
Is there any advantage to turning off fields in batches rather than individually?
Can the performance be improved by turning fields back on and off again?
These as well are directions for future work.

\section{Acknowledgements} \label{sec:acknowledgements}

We would like to thank the physics department at Michigan State University (MSU) for start-up funds that supported this work.
The numerical analysis was performed using computing resources provided by the Institute for Cyber-Enabled Research (ICER) at MSU.

\bibliography{Inh}

\begin{thebibliography}{64}
\bibitem [{sta()}]{starting_point_note}%
  \BibitemOpen
  \href@noop {} {}\bibinfo {note} {We begin at $s(0) = \tau(0) = 1/10$ rather than 0 for a purely technical reason. If the spins were to all begin in $\hat{\sigma}^x$ eigenstates, then under the mean-field equations, they would remain in $\hat{\sigma}^x$ eigenstates at all times and never approach the ground state of $H_0$ (regardless of $s(t)$ and $\tau(t)$). IQA would thus technically fail, but for reasons which have nothing to do with IQA. This peculiar behavior is specific to the $p$-spin model, however, and even just to the mean-field description of it, so we do not attach much significance to it --- we simply start at non-zero $s$ and $\tau$, where the ground state is not $\hat{\sigma}^x$ eigenstates, to circumvent the issue.}\BibitemShut {Stop}%
\bibitem [{cro()}]{crossing_average_note}%
  \BibitemOpen
  \href@noop {} {}\bibinfo {note} {For those realizations without any level crossing, the final energy approaches zero at large $T$ even under IQA, but since the fraction of such realizations quickly vanishes as $N$ increases, we find it cleaner to simply omit them from the average. We also observe in passing that even when there is no level crossing, the minimum gap under IQA is substantially smaller than under conventional QA, by at least an order of magnitude.}\BibitemShut {Stop}%
\end{thebibliography}
\appendix

\begin{widetext}

\section{Thermodynamics of the $p$-spin model under IQA} \label{sec:appendix1}

Here we describe the mean-field calculation for the ground state of the ferromagnetic $p$-spin model under IQA.
In fact, we first consider a more general situation, namely we calculate the free energy at inverse temperature $\beta$ for the Hamiltonian
\begin{equation} \label{eq:full_inhomogeneous_Hamiltonian}
H = -N s \Bigg( \frac{1}{N} \sum_{j=1}^N \hat{\sigma}_j^z \Bigg)^p - \sum_{j=1}^N \Gamma_j \hat{\sigma}_j^x,
\end{equation}
with arbitrary $\Gamma_j$.
The IQA results are then obtained by taking $\beta \rightarrow \infty$ and setting
\begin{equation} \label{eq:IQA_thermodynamic_fields}
\Gamma_j = \begin{cases} 1, \quad &0 \leq \tau < \frac{N-j}{N} \\ 0, \quad &\frac{N-j}{N} \leq \tau \leq 1 \end{cases}.
\end{equation}
Note that these calculations were already performed in Refs.~\cite{ExponentialSpeedupQuantum2018susa,QuantumAnnealing$p$spin2018susa} --- we include them here mainly to highlight the analogy with the subsequent dynamical calculations in App.~\ref{sec:appendix2} (also because the ground state is needed as the initial condition for the dynamics).

Starting from
\begin{equation} \label{eq:partition_function_definition}
Z \equiv \textrm{Tr} e^{-\beta H},
\end{equation}
we use the Suzuki-Trotter decomposition:
\begin{equation} \label{eq:partition_function_Suzuki_Trotter}
Z = \lim_{M \rightarrow \infty} \left( \exp{\left[ \frac{N \beta s}{M} \Bigg( \frac{1}{N} \sum_{j=1}^N \hat{\sigma}_j^z \Bigg)^p \right]} \exp{\left[ \frac{\beta}{M} \sum_{j=1}^N \Gamma_j \hat{\sigma}_j^x \right]} \right)^M.
\end{equation}
Insert a resolution of the identity in the $\hat{\sigma}^z$ basis between every factor.
Using $\sigma_j(k)$ without a hat to denote a classical variable taking values $\pm 1$, we have
\begin{equation} \label{eq:partition_function_classical}
Z = \lim_{M \rightarrow \infty} \sum_{\{ \sigma \}} \exp{\left[ \frac{N \beta s}{M} \sum_{k=1}^M \Bigg( \frac{1}{N} \sum_{j=1}^N \sigma_j(k) \Bigg)^p \right]} \prod_{k=1}^M \prod_{j=1}^N \big< \sigma_j(k) \big| \exp{\left[ \frac{\beta \Gamma_j}{M} \hat{\sigma}_j^x \right]} \big| \sigma_j(k-1) \big>,
\end{equation}
where the sum over $\{ \sigma \}$ denotes the $MN$ nested sums over $\sigma_j(k)$ for each $j$ and $k$.
Since the first exponential factor in Eq.~\eqref{eq:partition_function_classical} is a function solely of the quantities $m(k) \equiv N^{-1} \sum_j \sigma_j(k)$, we can write
\begin{equation} \label{eq:partition_function_order_parameters}
\begin{aligned}
Z = \lim_{M \rightarrow \infty} \sum_{\{ \sigma \}} \int_{-1}^1 \prod_{k=1}^M \textrm{d}m(k) \delta \bigg( m(k) - \frac{1}{N} \sum_{j=1}^N \sigma_j(k) \bigg) &\exp{\left[ \frac{N \beta s}{M} \sum_{k=1}^M m(k)^p \right]} \\
&\cdot \prod_{k=1}^M \prod_{j=1}^N \big< \sigma_j(k) \big| \exp{\left[ \frac{\beta \Gamma_j}{M} \hat{\sigma}_j^x \right]} \big| \sigma_j(k-1) \big>.
\end{aligned}
\end{equation}
Next express each $\delta$-function as the integral of a complex exponential:
\begin{equation} \label{eq:partition_function_delta_function_representation}
\delta \bigg( m(k) - \frac{1}{N} \sum_{j=1}^N \sigma_j(k) \bigg) = \int_{-i \infty}^{i \infty} \frac{N \beta \textrm{d}h(k)}{2\pi i M} \exp{\left[ -\frac{N \beta}{M} h(k) \bigg( m(k) - \frac{1}{N} \sum_{j=1}^N \sigma_j(k) \bigg) \right]}.
\end{equation}
After rearranging terms a bit, we thus have
\begin{equation} \label{eq:partition_function_path_integral_v1}
\begin{aligned}
Z &= \lim_{M \rightarrow \infty} \int_{-1}^1 \prod_{k=1}^M \textrm{d}m(k) \int_{-i \infty}^{i \infty} \prod_{k=1}^M \frac{N \beta \textrm{d}h(k)}{2\pi i M} \exp{\left[ \frac{N \beta s}{M} \sum_{k=1}^M m(k)^p - \frac{N \beta}{M} \sum_{k=1}^M h(k) m(k) \right]} \\
&\qquad \qquad \cdot \sum_{\{ \sigma \}} \exp{\left[ \frac{\beta}{M} \sum_{k=1}^M h(k) \sum_{j=1}^N \sigma_j(k) \right]} \prod_{k=1}^M \prod_{j=1}^N \big< \sigma_j(k) \big| \exp{\left[ \frac{\beta \Gamma_j}{M} \hat{\sigma}_j^x \right]} \big| \sigma_j(k-1) \big>.
\end{aligned}
\end{equation}
Note that the sum over $\{ \sigma \}$ now factors among different $j$.
In particular, define
\begin{equation} \label{eq:partition_function_single_spin_time_dependent}
Z_j[h] \equiv \sum_{\{ \sigma_j \}} \exp{\left[ \frac{\beta}{M} \sum_{k=1}^M h(k) \sigma_j(k) \right]} \prod_{k=1}^M \big< \sigma_j(k) \big| \exp{\left[ \frac{\beta \Gamma_j}{M} \hat{\sigma}_j^x \right]} \big| \sigma_j(k-1) \big>,
\end{equation}
which can be interpreted as the partition function for a single spin in the presence of (imaginary-time-dependent) longitudinal field $h(k)$ and transverse field $\Gamma_j$ (the sum over $\{ \sigma_j \}$ denotes the $M$ nested sums over $\sigma_j(k)$ for each $k$).
We have that
\begin{equation} \label{eq:partition_function_path_integral_v2}
Z = \lim_{M \rightarrow \infty} \int_{-1}^1 \prod_{k=1}^M \textrm{d}m(k) \int_{-i \infty}^{i \infty} \prod_{k=1}^M \frac{N \beta \textrm{d}h(k)}{2\pi i M} \exp{\left[ \frac{N \beta s}{M} \sum_{k=1}^M m(k)^p - \frac{N \beta}{M} \sum_{k=1}^M h(k) m(k) \right]} \prod_{j=1}^N Z_j[h].
\end{equation}

At large $N$, the integrals over $m(k)$ and $h(k)$ can be evaluated by saddle-point approximation.
Setting the derivatives of the exponent in Eq.~\eqref{eq:partition_function_path_integral_v2} to zero, we obtain the $2M$ saddle-point equations
\begin{equation} \label{eq:partition_function_saddle_point_equations}
ps m(k)^{p-1} = h(k), \qquad m(k) = \frac{M}{N \beta} \sum_{j=1}^N \frac{\partial \log{Z_j[h]}}{\partial h(k)}.
\end{equation}
Furthermore,
\begin{equation} \label{eq:partition_function_single_spin_expectation_value}
\frac{M}{\beta} \frac{\partial \log{Z_j[h]}}{\partial h(k)} = \sum_{\{ \sigma_j \}} \sigma_j(k) \frac{1}{Z_j[h]} \exp{\left[ \frac{\beta}{M} \sum_{k'=1}^M h(k') \sigma_j(k') \right]} \prod_{k'=1}^M \big< \sigma_j(k') \big| \exp{\left[ \frac{\beta \Gamma_j}{M} \hat{\sigma}_j^x \right]} \big| \sigma_j(k'-1) \big>,
\end{equation}
which can be interpreted as the thermal expectation value of $\sigma_j(k)$.
Once we solve these equations, the free energy density $f$ is given by
\begin{equation} \label{eq:partition_function_saddle_point_free_energy}
f \equiv -\frac{1}{N \beta} \log{Z} \sim -\frac{s}{M} \sum_{k=1}^M m(k)^p + \frac{1}{M} \sum_{k=1}^M h(k) m(k) - \frac{1}{N \beta} \sum_{j=1}^N \log{Z_j[h]},
\end{equation}
using the solution to Eq.~\eqref{eq:partition_function_saddle_point_equations} for $m(k)$ and $h(k)$.
In the event of multiple solutions, we use that which gives the lowest value of $f$.

For one thing, note that even though the integration contour for $h(k)$ is along the imaginary axis, the solution to the saddle-point equations will have real $h(k)$.
This simply means deforming the contour to pass through that real value, and so we will treat $h(k)$ as real from now on.
Also note that a consistent solution to the saddle-point equations has $m(k)$ and $h(k)$ be independent of $k$ --- this is clear for the left equation in Eq.~\eqref{eq:partition_function_saddle_point_equations}, and we will see it momentarily for the right equation as well.
While technically this does not rule out there also being non-constant solutions, we will assume that the constant solutions have lowest free energy (the ``static approximation''), and thus set $m(k) = m$ and $h(k) = h$.

Under the static approximation, $Z_j[h]$ simplifies immensely: we have
\begin{equation} \label{eq:partition_function_single_spin_static}
Z_j[h] = \sum_{\{ \sigma_j \}} \exp{\left[ \frac{\beta h}{M} \sum_{k=1}^M \sigma_j(k) \right]} \prod_{k=1}^M \big< \sigma_j(k) \big| \exp{\left[ \frac{\beta \Gamma_j}{M} \hat{\sigma}_j^x \right]} \big| \sigma_j(k-1) \big>,
\end{equation}
and note that this is precisely the expression that one would obtain when applying the Suzuki-Trotter decomposition to the partition function for a single spin in longitudinal field $h$ and transverse field $\Gamma_j$.
In other words,
\begin{equation} \label{eq:partition_function_single_spin_compact}
Z_j[h] = \textrm{Tr} e^{\beta h \hat{\sigma}_j^z + \beta \Gamma_j \hat{\sigma}_j^x},
\end{equation}
and the latter expression can be evaluated directly:
\begin{equation} \label{eq:partition_function_single_spin_result}
Z_j[h] = 2 \cosh{\beta \sqrt{h^2 + \Gamma_j^2}}.
\end{equation}
Applying the same reasoning to Eq.~\eqref{eq:partition_function_single_spin_expectation_value} gives
\begin{equation} \label{eq:partition_function_single_spin_expectation_compact}
\begin{aligned}
&\sum_{\{ \sigma_j \}} \sigma_j(k) \frac{1}{Z_j[h]} \exp{\left[ \frac{\beta h}{M} \sum_{k'=1}^M \sigma_j(k') \right]} \prod_{k'=1}^M \big< \sigma_j(k') \big| \exp{\left[ \frac{\beta \Gamma_j}{M} \hat{\sigma}_j^x \right]} \big| \sigma_j(k'-1) \big> \\
&\qquad \qquad \qquad \qquad = \frac{1}{Z_j[h]} \textrm{Tr} e^{(1 - \frac{k}{M}) (\beta h \hat{\sigma}_j^z + \beta \Gamma_j \hat{\sigma}_j^x)} \hat{\sigma}_j^z e^{\frac{k}{M} (\beta h \hat{\sigma}_j^z + \beta \Gamma_j \hat{\sigma}_j^x)}.
\end{aligned}
\end{equation}
By the cyclic property of the trace, this is indeed independent of $k$, and it amounts to the expectation value of $\hat{\sigma}_j^z$, which can also be evaluated directly:
\begin{equation} \label{eq:partition_function_single_spin_expectation_result}
\frac{1}{Z_j[h]} \textrm{Tr} \hat{\sigma}_j^z e^{\beta h \hat{\sigma}_j^z + \beta \Gamma_j \hat{\sigma}_j^x} = \frac{h}{\sqrt{h^2 + \Gamma_j^2}} \tanh{\beta \sqrt{h^2 + \Gamma_j^2}}.
\end{equation}
Thus the saddle-point equations reduce to
\begin{equation} \label{eq:partition_function_saddle_point_equations_static}
psm^{p-1} = h, \qquad m = \frac{1}{N} \sum_{j=1}^N \frac{h}{\sqrt{h^2 + \Gamma_j^2}} \tanh{\beta \sqrt{h^2 + \Gamma_j^2}},
\end{equation}
and the resulting free energy reduces to
\begin{equation} \label{eq:partition_function_free_energy_static}
f = -sm^p + hm - \frac{1}{N \beta} \sum_{j=1}^N \log{2 \cosh{\beta \sqrt{h^2 + \Gamma_j^2}}}.
\end{equation}

Now let us finally specialize to IQA.
Using Eq.~\eqref{eq:IQA_thermodynamic_fields} for $\Gamma_j$, the saddle-point equations become
\begin{equation} \label{eq:partition_function_saddle_point_equations_IQA_finite_T}
psm^{p-1} = h, \qquad m = (1 - \tau) \frac{h}{\sqrt{h^2 + 1}} \tanh{\beta \sqrt{h^2 + 1}} + \tau \tanh{\beta h},
\end{equation}
and the free energy density becomes
\begin{equation} \label{eq:partition_function_free_energy_IQA_finite_T}
f = -sm^p + hm - \frac{1-\tau}{\beta} \log{2 \cosh{\beta \sqrt{h^2 + 1}}} - \frac{\tau}{\beta} \log{2 \cosh{\beta h}}.
\end{equation}
Further taking $\beta \rightarrow \infty$, so that we are considering the ground state, the saddle-point equations become
\begin{equation} \label{eq:partition_function_saddle_point_equations_IQA_zero_T}
psm^{p-1} = h, \qquad m = (1 - \tau) \frac{h}{\sqrt{h^2 + 1}} + \tau \textrm{sgn}[h],
\end{equation}
and the free energy density (now the ground-state energy density) becomes
\begin{equation} \label{eq:partition_function_free_energy_IQA_zero_T}
f = -sm^p + hm - (1 - \tau) \sqrt{h^2 + 1} - \tau |h|.
\end{equation}

Technically we do not need the thermodynamics in order to carry out the dynamical calculations reported in the main text, but we do need the ground state at the starting values of $s$ and $\tau$ for the IQA protocol.
The above analysis shows that the partition function amounts to that of non-interacting spins in an additional longitudinal field $h$, albeit with the field determined self-consistently from the saddle-point equations.
By taking $\beta \rightarrow \infty$, this also holds for the ground state: once we solve the saddle-point equations for $h$, the ground state is simply the product state in which spin $j$ is in the ground state of the Hamiltonian $-h \hat{\sigma}_j^z - \Gamma_j \hat{\sigma}_j^x$.
This is what we use as initial conditions when studying the IQA dynamics.

\section{Dynamics of the $p$-spin model under IQA}\label{sec:appendix2}

Here we derive the mean-field equations given in the main text (Eqs.~\eqref{eq:mean_field_Schrodinger_equation} through~\eqref{eq:average_magnetization_definition}) and used to obtain the results reported there.
Note that the entire analysis is quite analogous to that in App.~\ref{sec:appendix1}, simply using a Keldysh path integral in place of the thermodynamic path integral.
Once again, we actually consider the dynamics under the more general Hamiltonian
\begin{equation} \label{eq:full_inhomogeneous_time_dependent_Hamiltonian}
H(t) = -N s(t) \Bigg( \frac{1}{N} \sum_{j=1}^N \hat{\sigma}_j^z \Bigg)^p - \sum_{j=1}^N \Gamma_j(t) \hat{\sigma}_j^x,
\end{equation}
where $s(t)$ and each $\Gamma_j(t)$ are arbitrary functions of time --- IQA is a special case.
Denote the initial state of the system by the density matrix $\hat{\rho}(0)$.
We will later take $\hat{\rho}(0)$ to be the product state obtained from the thermodynamic calculation (see App.~\ref{sec:appendix1}), but for now we keep it general.

As discussed in the main text, we are particularly interested in the time evolution of the magnetization $m^z(t) \equiv N^{-1} \sum_j \langle \hat{\sigma}_j^z(t) \rangle$.
Note that all expectation values of $\hat{\sigma}^z$ operators at any combination of times can be obtained from derivatives of the generating functional
\begin{equation} \label{eq:Keldysh_generating_functional}
\mathcal{Z} \equiv \textrm{Tr} \Big( \mathcal{T} e^{-i \int_0^T \textrm{d}t \big[ H(t) - \sum_j \xi_j^+(t) \hat{\sigma}_j^z \big]} \Big) \hat{\rho}(0) \Big( \mathcal{T} e^{-i \int_0^T \textrm{d}t \big[ H(t) - \sum_j \xi_j^-(t) \hat{\sigma}_j^z \big]} \Big)^{\dag},
\end{equation}
where $\mathcal{T}$ indicates that the exponential is time-ordered.
For example,
\begin{equation} \label{eq:Keldysh_derivative_example_1}
-i \frac{\delta \mathcal{Z}}{\delta \xi_j^+(t)} \Bigg|_{\xi^+ = \xi^- = 0} = \textrm{Tr} \hat{\sigma}_j^z \Big( \mathcal{T} e^{-i \int_0^t \textrm{d}s H(s)} \Big) \hat{\rho}(0) \Big( \mathcal{T} e^{-i \int_0^t \textrm{d}s H(s)} \Big)^{\dag} = \big< \hat{\sigma}_j^z(t) \big>,
\end{equation}
\begin{equation} \label{eq:Keldysh_derivative_example_2}
\frac{\delta^2 \mathcal{Z}}{\delta \xi_j^+(t) \delta \xi_j^-(t')} \Bigg|_{\xi^+ = \xi^- = 0} = \textrm{Tr} \hat{\sigma}_j^z \Big( \mathcal{T} e^{-i \int_0^t \textrm{d}s H(s)} \Big) \hat{\rho}(0) \Big( \mathcal{T} e^{-i \int_0^{t'} \textrm{d}s H(s)} \Big)^{\dag} \hat{\sigma}_j^z \Big( \mathcal{T} e^{-i \int_{t'}^t \textrm{d}s H(s)} \Big)^{\dag} = \big< \hat{\sigma}_j^z(t') \hat{\sigma}_j^z(t) \big>.
\end{equation}
The generating functional $\mathcal{Z}$ is thus the dynamical analogue of the partition function $Z$.
Note that it is useful to have separate ``source terms'' $\xi^+$ and $\xi^-$ to control the time-ordering of expectation values (e.g., Eq.~\eqref{eq:Keldysh_derivative_example_2} has $\hat{\sigma}_j^z(t')$ to the left of $\hat{\sigma}_j^z(t)$ even though $t' < t$).

To begin, use the Suzuki-Trotter decomposition on each time-ordered exponential (this is in fact the definition of the time-ordered exponential): divide the interval $[0, T]$ into $M$ timesteps, denoting $\Delta t \equiv T/M$ and $t_k \equiv k \Delta t$, and then
\begin{equation} \label{eq:Keldysh_Suzuki_Trotter}
\begin{aligned}
&\mathcal{T} e^{-i \int_0^T \textrm{d}t \big[ H(t) - \sum_j \xi_j^+(t) \hat{\sigma}_j^z \big]} \\
&\qquad \qquad = \lim_{M \rightarrow \infty} \exp{\left[ i \Delta t N s(t_M) \Bigg( \frac{1}{N} \sum_{j=1}^N \hat{\sigma}_j^z \Bigg)^p + i \Delta t \sum_{j=1}^N \xi_j^+(t_M) \hat{\sigma}_j^z \right]} \exp{\left[ i \Delta t \sum_{j=1}^N \Gamma_j(t_M) \hat{\sigma}_j^x \right]} \\
&\qquad \qquad \qquad \qquad \qquad \cdots \exp{\left[ i \Delta t N s(t_1) \Bigg( \frac{1}{N} \sum_{j=1}^N \hat{\sigma}_j^z \Bigg)^p + i \Delta t \sum_{j=1}^N \xi_j^+(t_1) \hat{\sigma}_j^z \right]} \exp{\left[ i \Delta t \sum_{j=1}^N \Gamma_j(t_1) \hat{\sigma}_j^x \right]}.
\end{aligned}
\end{equation}
Again insert a resolution of the identity in the $\hat{\sigma}^z$ basis between every factor (for each of the two exponentials in Eq.~\eqref{eq:Keldysh_generating_functional}).
The generating functional then becomes a sum over $2MN$ classical Ising variables $\sigma_j^{\pm}(k)$:
\begin{equation} \label{eq:Keldysh_path_integral}
\begin{aligned}
\mathcal{Z} &= \lim_{M \rightarrow \infty} \sum_{\{ \sigma \}} \exp{\left[ i \Delta t N \sum_{k=1}^M s(t_k) \left( \Bigg( \frac{1}{N} \sum_{j=1}^N \sigma_j^+(k) \Bigg)^p - \Bigg( \frac{1}{N} \sum_{j=1}^N \sigma_j^-(k) \Bigg)^p \right) \right]} \\
&\qquad \qquad \qquad \cdot \exp{\left[ i \Delta t \sum_{k=1}^M \sum_{j=1}^N \Big( \xi_j^+(t_k) \sigma_j^+(k) - \xi_j^-(t_k) \sigma_j^-(k) \Big) \right]} \\
&\qquad \qquad \qquad \qquad \cdot \prod_{k=1}^M \prod_{j=1}^N \big< \sigma_j^+(k) \big| e^{i \Delta t \Gamma_j(t_k) \hat{\sigma}_j^x} \big| \sigma_j^+(k-1) \big> \big< \sigma_j^-(k-1) \big| e^{-i \Delta t \Gamma_j(t_k) \hat{\sigma}_j^x} \big| \sigma_j^-(k) \big> \\
&\qquad \qquad \qquad \qquad \qquad \cdot \big< \sigma_1^+(0), \cdots, \sigma_N^+(0) \big| \hat{\rho}(0) \big| \sigma_1^-(0), \cdots, \sigma_N^-(0) \big>.
\end{aligned}
\end{equation}
Introduce $\delta$-functions setting $m^{\pm}(t_k) = N^{-1} \sum_j \sigma_j^{\pm}(k)$ (compare to Eq.~\eqref{eq:partition_function_order_parameters}):
\begin{equation} \label{eq:Keldysh_order_parameters}
\begin{aligned}
\mathcal{Z} &= \lim_{M \rightarrow \infty} \sum_{\{ \sigma \}} \int_{-1}^1 \prod_{k=1}^M \textrm{d}m^+(t_k) \textrm{d}m^-(t_k) \delta \bigg( m^+(t_k) - \frac{1}{N} \sum_{j=1}^N \sigma_j^+(k) \bigg) \delta \bigg( m^-(t_k) - \frac{1}{N} \sum_{j=1}^N \sigma_j^-(k) \bigg) \\
&\qquad \qquad \cdot \exp{\left[ i \Delta t N \sum_{k=1}^M s(t_k) \Big( m^+(t_k)^p - m^-(t_k)^p \Big) + i \Delta t \sum_{k=1}^M \sum_{j=1}^N \Big( \xi_j^+(t_k) \sigma_j^+(k) - \xi_j^-(t_k) \sigma_j^-(k) \Big) \right]} \\
&\qquad \qquad \qquad \qquad \cdot \prod_{k=1}^M \prod_{j=1}^N \big< \sigma_j^+(k) \big| e^{i \Delta t \Gamma_j(t_k) \hat{\sigma}_j^x} \big| \sigma_j^+(k-1) \big> \big< \sigma_j^-(k-1) \big| e^{-i \Delta t \Gamma_j(t_k) \hat{\sigma}_j^x} \big| \sigma_j^-(k) \big> \\
&\qquad \qquad \qquad \qquad \qquad \cdot \big< \sigma_1^+(0), \cdots, \sigma_N^+(0) \big| \hat{\rho}(0) \big| \sigma_1^-(0), \cdots, \sigma_N^-(0) \big>,
\end{aligned}
\end{equation}
and then express each $\delta$-function as the integral of a complex exponential:
\begin{equation} \label{eq:Keldysh_expanded_path_integral}
\begin{aligned}
\mathcal{Z} &= \lim_{M \rightarrow \infty} \int_{-1}^1 \prod_{k=1}^M \textrm{d}m^+(t_k) \textrm{d}m^-(t_k) \int_{-\infty}^{\infty} \prod_{k=1}^M \frac{N \Delta t \textrm{d}h^+(t_k)}{2\pi} \frac{N \Delta t \textrm{d}h^-(t_k)}{2\pi} \\
&\qquad \qquad \cdot \exp{\left[ i \Delta t N \sum_{k=1}^M s(t_k) \Big( m^+(t_k)^p - m^-(t_k)^p \Big) - i \Delta t N \sum_{k=1}^M \Big( h^+(t_k) m^+(t_k) - h^-(t_k) m^-(t_k) \Big) \right]} \\
&\qquad \qquad \qquad \cdot \sum_{\{ \sigma \}} \exp{\left[ i \Delta t \sum_{k=1}^M \sum_{j=1}^N \Big( \big[ h^+(t_k) + \xi_j^+(t_k) \big] \sigma_j^+(k) - \big[ h^-(t_k) + \xi_j^-(t_k) \big] \sigma_j^-(k) \Big) \right]} \\
&\qquad \qquad \qquad \qquad \cdot \prod_{k=1}^M \prod_{j=1}^N \big< \sigma_j^+(k) \big| e^{i \Delta t \Gamma_j(t_k) \hat{\sigma}_j^x} \big| \sigma_j^+(k-1) \big> \big< \sigma_j^-(k-1) \big| e^{-i \Delta t \Gamma_j(t_k) \hat{\sigma}_j^x} \big| \sigma_j^-(k) \big> \\
&\qquad \qquad \qquad \qquad \qquad \cdot \big< \sigma_1^+(0), \cdots, \sigma_N^+(0) \big| \hat{\rho}(0) \big| \sigma_1^-(0), \cdots, \sigma_N^-(0) \big>.
\end{aligned}
\end{equation}

Assume that the initial state is a product state, i.e., $\hat{\rho}(0) = \bigotimes_j \hat{\rho}_j(0)$.
We can then begin to simplify this expression.
The sum over $\{ \sigma \}$ factors among different $j$, so define
\begin{equation} \label{eq:Keldysh_effective_generating_functional}
\begin{aligned}
\mathcal{Z}_j[h] &\equiv \sum_{\{ \sigma_j \}} \exp{\left[ i \Delta t \sum_{k=1}^M \Big( \big[ h^+(t_k) + \xi_j^+(t_k) \big] \sigma_j^+(k) - \big[ h^-(t_k) + \xi_j^-(t_k) \big] \sigma_j^-(k) \Big) \right]} \\
&\qquad \qquad \cdot \left( \prod_{k=1}^M \big< \sigma_j^+(k) \big| e^{i \Delta t \Gamma_j(t_k) \hat{\sigma}_j^x} \big| \sigma_j^+(k-1) \big> \big< \sigma_j^-(k-1) \big| e^{-i \Delta t \Gamma_j(t_k) \hat{\sigma}_j^x} \big| \sigma_j^-(k) \big> \right) \big< \sigma_j^+(0) \big| \hat{\rho}_j(0) \big| \sigma_j^-(0) \big>.
\end{aligned}
\end{equation}
The generating functional can then be written as
\begin{equation} \label{eq:Keldysh_generating_functional_compact}
\begin{aligned}
\mathcal{Z} &= \lim_{M \rightarrow \infty} \int_{-1}^1 \prod_{k=1}^M \textrm{d}m^+(t_k) \textrm{d}m^-(t_k) \int_{-\infty}^{\infty} \prod_{k=1}^M \frac{N \Delta t \textrm{d}h^+(t_k)}{2\pi} \frac{N \Delta t \textrm{d}h^-(t_k)}{2\pi} \\
&\qquad \cdot \exp{\left[ i \Delta t N \sum_{k=1}^M s(t_k) \Big( m^+(t_k)^p - m^-(t_k)^p \Big) - i \Delta t N \sum_{k=1}^M \Big( h^+(t_k) m^+(t_k) - h^-(t_k) m^-(t_k) \Big) \right]} \prod_{j=1}^N \mathcal{Z}_j[h].
\end{aligned}
\end{equation}
At large $N$, we evaluate the integrals over $m^{\pm}(t_k)$ and $h^{\pm}(t_k)$ via saddle-point approximation.
This gives the saddle-point equations
\begin{equation} \label{eq:Keldysh_saddle_point_equations_initial}
\begin{aligned}
p s(t_k) m^+(t_k)^{p-1} &= h^+(t_k), \qquad &m^+(t_k) &= \frac{1}{i \Delta t N} \sum_{j=1}^N \frac{\partial \log{Z_j[h]}}{\partial h^+(t)}, \\
p s(t_k) m^-(t_k)^{p-1} &= h^-(t_k), \qquad &m^-(t_k) &= \frac{1}{i \Delta t N} \sum_{j=1}^N \frac{\partial \log{Z_j[h]}}{\partial h^-(t)}.
\end{aligned}
\end{equation}
A consistent solution to these equations has $m^+(t_k) = m^-(t_k) = m(t_k)$ and $h^+(t_k) = h^-(t_k) = h(t_k)$.
Note that in this case, $\mathcal{Z}_j[h]$ is quite literally the generating functional for a single spin $j$ in longitudinal field $h(t)$ and transverse field $\Gamma_j(t)$:
\begin{equation} \label{eq:Keldysh_effective_generating_functional_continuum}
\mathcal{Z}_j[h] \sim \textrm{Tr} \Big( \mathcal{T} e^{i \int_0^T \textrm{d}t \big[ \big( h(t) + \xi_j^+(t) \big) \hat{\sigma}_j^z + \Gamma_j(t) \hat{\sigma}_j^x \big]} \Big) \hat{\rho}_j(0) \Big( \mathcal{T} e^{i \int_0^T \textrm{d}t \big[ \big( h(t) + \xi_j^-(t) \big) \hat{\sigma}_j^z + \Gamma_j(t) \hat{\sigma}_j^x \big]} \Big)^{\dag}.
\end{equation}
Furthermore, differentiating $\mathcal{Z}_j[h]$ with respect to $h^{\pm}(t)$ is equivalent to differentiating with respect to $\xi_j^{\pm}(t)$, and according to Eq.~\eqref{eq:Keldysh_derivative_example_1},
\begin{equation} \label{eq:Keldysh_single_spin_derivative}
-i \frac{\delta \mathcal{Z}_j}{\delta \xi_j^+(t)} \Bigg|_{\xi^+ = \xi^- = 0} = i \frac{\delta \mathcal{Z}_j}{\delta \xi_j^-(t)} \Bigg|_{\xi^+ = \xi^- = 0} = \big< \hat{\sigma}_j^z(t) \big>_{\textrm{eff}}.
\end{equation}
where $\langle \, \cdot \, \rangle_{\textrm{eff}}$ denotes the expectation value using the \textit{non-interacting} Hamiltonian $-h(t) \hat{\sigma}_j^z - \Gamma_j(t) \hat{\sigma}_j^x$.
Lastly, note that $\mathcal{Z}_j[h]|_{\xi^+ = \xi^- = 0} = 1$ regardless of $h$, and thus the derivative of $\log{\mathcal{Z}_j[h]}$ equals the derivative of $\mathcal{Z}_j[h]$ itself (at least when the source terms are set to zero).
Putting all of this together, the saddle-point equations amount to
\begin{equation} \label{eq:Keldysh_saddle_point_equations}
p s(t) m(t)^{p-1} = h(t), \qquad m(t) = \frac{1}{N} \sum_{j=1}^N \big< \hat{\sigma}_j^z(t) \big>_{\textrm{eff}}.
\end{equation}

The order parameter $m(t)$ is nothing more than the magnetization at time $t$.
To see this formally, note that the average magnetization at $t_k$ is obtained by inserting a factor of $N^{-1} \sum_j \sigma_j^{\pm}(k)$ into the generating functional (either sign is allowed).
This becomes $m^{\pm}(t)$ upon introducing $\delta$-functions, and it is given by the saddle-point value upon making a saddle-point approximation.
In other words, the average magnetization equals the location of the saddle point in $m^{\pm}(t)$.

The form of Eq.~\eqref{eq:Keldysh_saddle_point_equations} suggests an efficient recursive method to determine the magnetization as a function of time.
Pick a small timestep $\Delta t$, and assume $| \psi_j(t_k) \rangle$ is known for each $j$.
Use the following to determine $| \psi_j(t_{k+1}) \rangle$:
\begin{itemize}
\item Calculate $m(t_k) \equiv N^{-1} \sum_j \langle \hat{\sigma}_j^z(t_k) \rangle$.
\item Determine $h(t_k) \equiv p s(t_k) m(t_k)^{p-1}$.
\item Form $H_{\textrm{eff},j}(t_k) \equiv -h(t_k) \hat{\sigma}_j^z - \Gamma_j(t_k) \hat{\sigma}_j^x$.
\item Compute $| \psi_j(t_{k+1}) \rangle \approx \exp{[-i H_{\textrm{eff},j}(t_k) \Delta t]} | \psi_j(t_k) \rangle$ (separately for each $j$).
\end{itemize}
Recall that we begin by setting $| \psi_j(0) \rangle$ to be the ground state obtained from the thermodynamic calculation.
We then follow this procedure to obtain the results given in the main text.

\end{widetext}

\end{document}